\documentclass[12pt]{iopart}

\usepackage{iopams}
\usepackage{graphicx}
\usepackage{wrapfig}
\usepackage[labelfont=bf]{caption}
\usepackage{color}
\begin{document}

\title[He$^+_2$ electron-impact dissociation cross sections]{Electron-impact dissociation cross sections of vibrationally excited He$^+_2$ molecular ion}

\author{R. Celiberto$^{1,2}$, K.L. Baluja$^3$, R.K. Janev$^{4}$, V. Laporta$^{5,6,2}$}
\address{$^1$Dipartimento di Ingegneria Civile, Ambientale, del Territorio, Edile e di Chimica, Politecnico di Bari, Italy}
\address{$^2$Istituto di Nanotecnologie, CNR, Bari, Italy}
\address{$^3$Department of Physics and Astrophysics, University of Delhi, Delhi 110007, India}
\address{$^4$Institute of  Energy and Climate Research - Plasma Physics, Forschungszentrum J\"{u}lich GmbH Association EURATOM-FZJ, Partner in Trilateral Euregio Cluster, 52425 J\"{u}lich, Germany}
\address{$^5$Ohio Aerospace Institute, Dayton 45431, Ohio}
\address{$^6$Department of Physics and Astronomy, University College London, London WC1E 6BT, UK}
\ead{roberto.celiberto@poliba.it}
\ead{kl\_baluja@yahoo.com}
\ead{r.janev@fz-juelich.de}
\ead{vincenzo.laporta@nanotec.cnr.it}

\vspace{10pt}
\begin{indented}
\item[]June 2015
\end{indented}

\begin{abstract}
Electron-impact cross sections for the dissociation process of vibrationally excited He$_2^+$ molecular ion, as a function of the incident electron energy are calculated for the dissociative transition $\textrm{X}\,^2\Sigma_u^+\rightarrow \textrm{A}\,^2\Sigma_g^+$ by using the $R$-matrix method in the adiabatic-nuclei approximation. The potential energy curves for the involved electronic states and target properties, also calculated with the $R$-matrix method, were found to be in good agreement with the results reported in literature.
\end{abstract}

%
%
%
%
%


\section{Introduction}
He$_2^+$ is a simple molecular ion that is present in many natural and technological systems. It plays a dominant role in electric discharges~\cite{Katusi_et_al} and stellar medium~\cite{Stencil}, as well as in the chemistry of the early universe~\cite{Dalgarno, Mihajlov_et_al} and thermonuclear fusion~\cite{IAEA_summary_report}. In helium containing plasmas, electron-impact dissociative recombination and direct dissociation are among the main reactive processes of the He$_2^+$ ion leading to atomic and ionic species. Dissociative recombination is a low-energy process occurring through temporarily neutralization of He$_2^+$ ion followed by dissociation into two He atoms. It has been widely studied due its importance in relatively low-temperature system (see Ref.~\cite{Royal_Orel} and references therein). On the opposite side, direct dissociation of He$_2^+$ induced by electron-impact into He atom and He$^+$ ion requires fast electrons with a kinetic energy equal or larger than the excitation energy of the process, which amounts, at the ground state equilibrium bond-length, to about $10$~eV. Corresponding temperatures of this order of magnitude of energies are typical for the divertor plasmas of toroidal fusion devices. The plasma cooling in the divertor region favors the recombination processes which allow for the formation of molecular excited species, so that plasma chemistry, involving gas-phase and surface becomes rather complex.

In the present work we undertake a theoretical study of cross sections for dissociative excitation induced by electron-impact which, to the best of our knowledge, has not been studied before theoretically or experimentally. The process can be represented as,
\begin{equation}
e + \textrm{He}_2^+(\textrm{X}\,^2\Sigma_u^+,v=0-23) \rightarrow e + \mathrm{He}_2^ + (\textrm{A}\,^2\Sigma^+_g) \rightarrow e + \mathrm{He}+\mathrm{He^+}\,, \label{process}
\end{equation}
where $v$ represent the vibrational level of He$_2^+$. The molecular states $\textrm{X}\,^2\Sigma_u^+$ and $\textrm{A}\,^2\Sigma^+_g$ are degenerate at infinitely large internuclear distances $R$ and the energy splitting of these two states at finite $R$ is a result of the resonant one-electron exchange between the two ion cores He$^+$.

The cross sections, calculated for all the 24 vibrational levels of the ground state potential well, were obtained by using the $R$-matrix method~\cite{Tennyson} in connection with the \textit{adiabatic nuclei} (AN) approximation~\cite{Lane, Hazi}. The $R$-matrix method, in fact, assumes an excitation of the target occurring at a fixed internuclear distance, at which the scattering $T$-matrix is calculated, so that the nuclear motion is suppressed and a purely electronic excitation cross section is produced. In order to take into account the vibrational dynamics of the process (\ref{process}) in a more realistic way, we used in our calculations the well-known AN approximation, which is based on a vibrational average of the scattering $T$-matrix calculated for a sufficient number of internuclear distances.

The paper is organized as follows: In Section~\ref{sec:Rmatrix} the $R$-matrix method is outlined and the construction of target states within the fixed-nuclei approximation is discussed. In Section~\ref{sec:adiabatic} the main equations of AN approximation are formulated and in Section~\ref{sec:results} we present and discuss our results. Section~\ref{sec:conclusions}, finally, concludes the work.

\section{$R$-matrix method \label{sec:Rmatrix}}
The main facet of the $R$-matrix method is the division of configurational space into an inner and an outer region. The inner region is defined as the volume of a sphere of radius $a$ centered at the center-of-mass of the target molecule. In this work we have set $a = 12$~a.u. This region is constructed so that the wave functions of all $N$ except the scattering electron vanish at boundary of the sphere. In this region the exchange effects, \emph{i.e.} the short range electron-electron correlations and polarization effects, are important. Implicitly this method assumes that the Pauli principle, which asserts that all inner region electrons are identical and any many electron wave function must be anti-symmetric to interchange of these electrons. In the outer region it is assumed that one electron can be considered to be distinct. This electron therefore moves in a local potential arising from its long-range interaction with the target. The exchange effects are neglected in this region. The construction of matrix $R$ provides the link between the inner and outer region. The $R$-matrix method has some distinct advantages. The main advantage is that the inner region problem needs to be solved only once.  The energy dependence is obtained entirely from the solution of the much simpler outer region case. This allows us to generate solutions at a large number of energies at a minimal extra computational effort. This generates cross sections on a fine energy grid that helps in analyzing the nature of resonances that may arise.

In the inner region the total wave function is expanded in a configuration-interaction (CI) basis which takes the following form for each total orbital angular momentum, spin and parity combination:
\begin{equation}
\Psi_k^{N+1} = {\cal A}\,\sum_{i,j} \Phi_i^N(\textbf{x}_1\ldots\textbf{x}_N)\, \xi_j (\textbf{x}_{N+1})\,a_{ijk} \,+\, \sum_m\chi_m(\textbf{x}_1\ldots\textbf{x}_{N+1})\,b_{mk}\,,	\label{RM wave function}
\end{equation}
where $\cal A$ is an anti-symmetrization operator, $\textbf{x}_N$ are the spatial and spin coordinates of the $N$-th electron, $\Phi_i^N$ is the wave function of the $i$-th target state, $\xi_j$ are the continuum orbitals of the scattering electron, $k$ represents a particular $R$-matrix basis function. The variational coefficients $a_{ijk}$ and $b_{mk}$ are determined by matrix diagonalization. Furthermore the electrons, whose space-spin coordinates are represent by $\textbf{x}_i$, must obey the Pauli principle and are therefore anti-symmetrized by operator $\cal A$. In practical implementations for generating the configurations which make up this term it is often necessary to impose a constraint on the coupling of the first $N$ electrons to ensure that the target wave function does not get contaminated by states with the same configuration but different space-spin symmetry. The second summation involves configurations which have no amplitude on the $R$-matrix boundary and where all electrons are placed in orbitals associated with the target. Since they are confined to a finite volume of space they will be referred to as $L^2$ configurations. Such configurations are essential to relax the constraint of in orthogonalization between the continuum orbitals and those belonging to the target of the same symmetry. In more sophisticated models the $L^2$ configurations are also used to model the effects of target polarization.  The remaining information that is required to set-up the outer region problem concerns properties of the target. The target state energies relative to the ground state are needed as they give the energies of the asymptotic channels. The multipole moments associated with these target states are also required as they determine the outer region. We include only the dipole and the quadrupole transition moments in the present work.

In our calculations we have 16 scattering symmetries. These are singlet and triplet spin states of the 8 irreducible representations of $D_{2h}$ point group. We have included up to $g$ partial waves contribution to the cross sections, the effect of the remaining partial waves is included via a Coulomb-Born closure approximation. The $R$-matrix constructed at 12~a.u. is propagated up to 50~a.u. where the effect of multipole moments is negligible~\cite{Baluja_et_al}. The scattering solutions are then matched with the Coulomb boundary conditions to extract $K$ matrices, $T$ matrices and other observables. We have employed UKRMol code suite of programs~\cite{RM_code}.

\subsection{Target states}
At the $R$-matrix boundary between the inner and outer regions, the amplitude of the target orbitals must be negligible. This restricts the use of extended diffuse functions in a chosen basis set. Also a single basis set is mandatory to represent all the target states included in the calculation.

The basis set employed in this work is the cc-pVTZ Gaussian basis set for He$_2^+$ molecule. This set includes polarization functions. The electronic configuration for the ground state $\textrm{X}\,^2\Sigma_u^+$ of He$_2^+$ in its natural symmetry is ($1\sigma_g)^2$ ($1\sigma_u$). The molecule is treated in a reduced $D_{2h}$ symmetry in which there are eight symmetries: $A_g$, $A_u$, $B_{1g}$, $B_{1u}$, $B_{2g}$, $B_{3g}$, $B_{2u}$, $B_{3u}$. The calculations were performed at the different equilibrium bond-lengths required in the calculation of AN cross sections. In particular for the equilibrium distance of 2.042~a.u. the self-consistent field (SCF) energy for the X state is $-4.92150$~a.u. In the close coupling expansion of the trial wave function of the scattering system, we included the X state and the first excited state $\textrm{A}\,^2\Sigma_g^+$, $(1\sigma_g)^2$ $(2\sigma_g)$. CI wave functions are used to represent each target state. In our CI model, we have the occupied orbitals which are augmented by the virtual (vacant) molecular orbitals. We include virtual orbitals up to $7a_g,\ 3b_{2u},\ 3b_{3u},\ 1b_{1g},\ 7b_{1u},\ 3b_{2g},\ 3b_{3g},\ 1a_u$. The three molecular orbitals were free to move into the entire active space. The vertical excitation energy of the A state at equilibrium distance is 9.954 eV with respect to the X state. We obtained for the dipole transition moment of X-A transition a value of 0.96856~a.u.
\begin{figure}
\centering
\includegraphics[width=5cm, angle=-90]{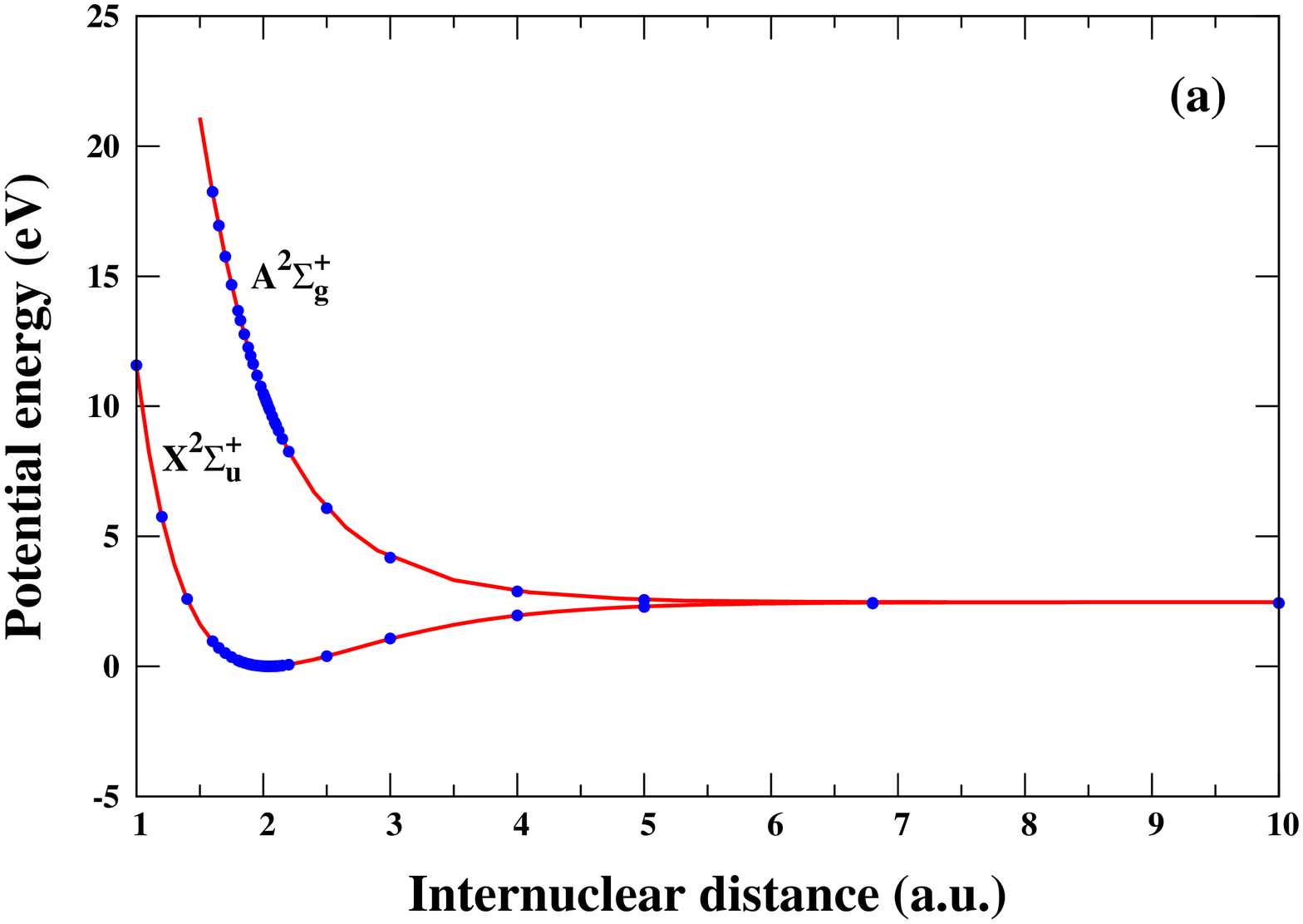}
\includegraphics[width=5cm, angle=-90]{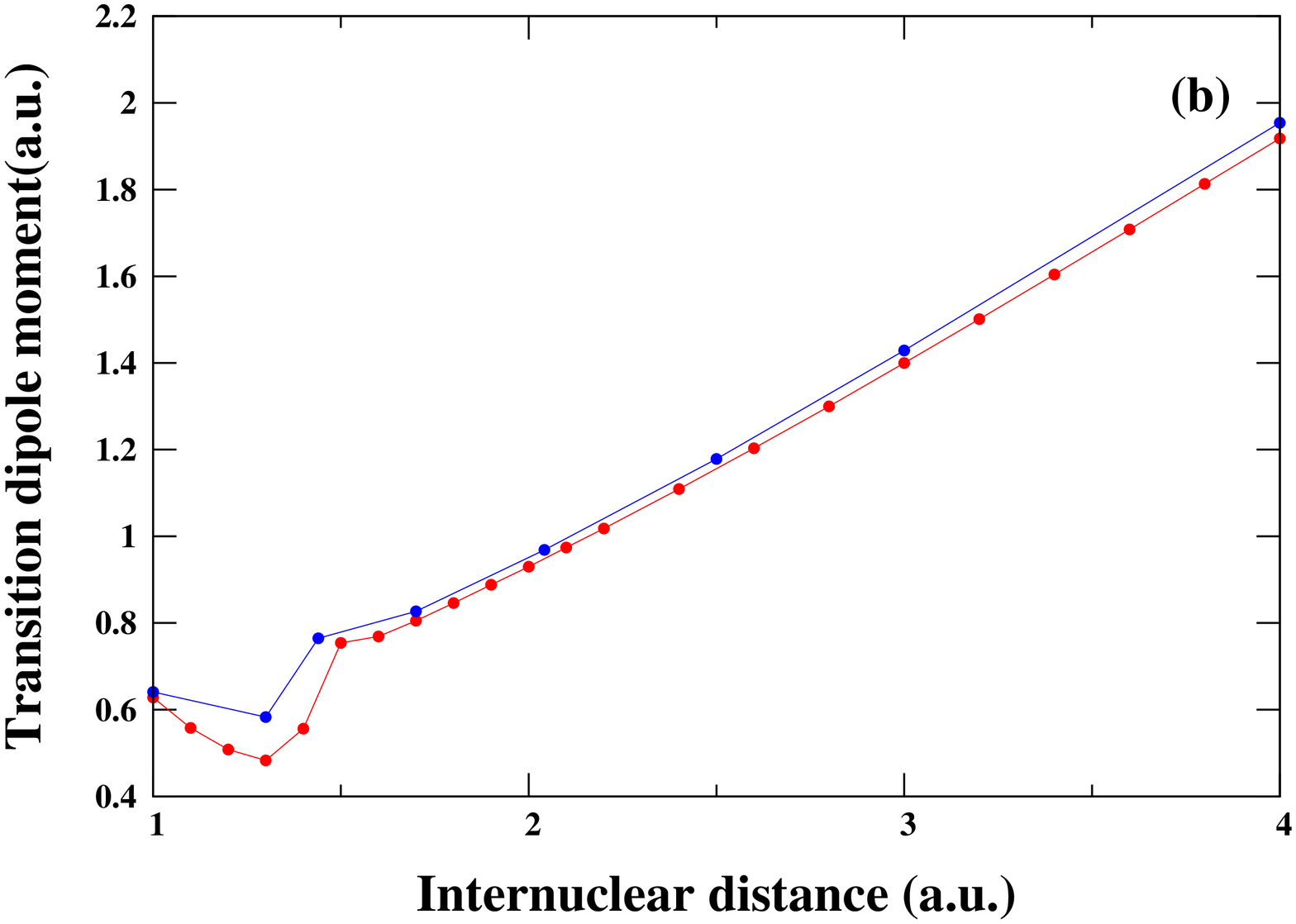}
\caption{\small ($a$) Ground ($\textrm{X}\,^2\Sigma_u^+$) and excited ($\textrm{A}\,^2\Sigma_g^+$) $R$-matrix electronic potential energies and ($b$) transition dipole moments compared with the calculations taken from Refs.~\cite{Xie_et_al} and \cite{Metropoulos_et_al} respectively. \label{fig:potentials}}
\end{figure}

Figures~\ref{fig:potentials}($a$) and ($b$) show a comparison of our $R$-matrix potential curves and transition dipole moments, with those of Refs.~\cite{Xie_et_al} and \cite{Metropoulos_et_al} respectively. The agreement is quite good in both cases. The ground and excited states of He$_2^+$ support respectively 24 and 2 vibrational levels. The corresponding eigenvalues, calculated by using the curves of Ref.~\cite{Xie_et_al}, are shown in Table~\ref{table:Eigenvalues}. The shallow minimum of A state appears at $R\approx 8.7$~a.u. an internuclear distance far away from the Franck-Condon (FC) regions of $\textrm{X}\,^2\Sigma_u^+$ so it cannot affect the dissociation process in any way.
\begin{table}
\begin{center}
\begin{tabular}{|c|c|c|c|c|}
\hline
\centering
~~~~$v$~~~~&$\textrm{X}\,^2\Sigma_u^+$  & $\textrm{X}\,^2\Sigma_u^+$        &$\textrm{A}\,^2\Sigma_g^+$  & $\textrm{A}\,^2\Sigma_g^+$\\
        & Present results & Ref. \cite{Tung_et_al} & Present results & Ref. \cite{Xie_et_al}\\
\hline
0	&	0.1042	&	0.1042	&	2.4733 & 2.4730\\
1	&	0.3069	&	0.3061	&	2.4743 & 2.4740\\
2	&	0.4997	&	0.4992    &        &\\		
3	&	0.6839	&	0.6837    &        &\\		
4	&	0.8597	&	0.8594    &        &\\		
5	&	1.0265	&	1.0264     &        &\\		
6	&	1.1846	&	1.1846     &        &\\		
7	&	1.3346	&	1.3340     &        &\\		
8	&	1.4761	&	1.4745     &        &\\		
9	&	1.6094	&	1.6062     &        &\\		
10	&	1.7347	&	1.7289     &        &\\		
11	&	1.8521	&	1.8426     &        &\\		
12	&	1.9607	&	1.9472     &        &\\		
13	&	2.0597	&	2.0426     &        &\\		
14	&	2.1485	&	2.1288     &        &\\		
15	&	2.2266	&	2.2056     &        &\\		
16	&	2.2938	&	2.2729     &        &\\		
17	&	2.3501	&	2.3306     &        &\\		
18	&	2.3952	&	2.3785     &        &\\		
19	&	2.4292	&	2.4165     &        &\\		
20	&	2.4523	&	2.4446     &        &\\		
21	&	2.4651	&	2.4627     &        &\\		
22	&	2.4700	&	2.4715     &        &\\		
23	&	2.4740	&	2.4739     &        &\\		
\hline
\end{tabular}
\caption{\label{table:Eigenvalues}\small Vibrational eigenvalues (eV) for the $\textrm{X}\,^2\Sigma_u^+$ and $\textrm{A}\,^2\Sigma_g^+$ electronic states compared with those from Ref.~\cite{Tung_et_al} and  Ref. \cite{Xie_et_al} respectively. All the values are referred to the bottom of the ground state potential curve.}
\end{center}
\end{table}

\section{Adiabatic nuclei approximation \label{sec:adiabatic}}
The AN approximation is well-known in scattering theory and the interested reader can refer to the rich literature on the subject~\cite{Lane, Hazi} for details. We will only recall here the main aspects of its formulation for diatomic molecules and state the relevant equations useful for our discussion.

The AN in its standard treatment, is based on the decoupling of the nuclear and electronic motion, so that the total scattering wave function can be approximated by the product of the electronic wave function in the fixed-nuclei approximation, \textit{i.e.} parametrically dependent on the internuclear distance $R$, and the nuclear ro-vibrational wave function. This factorization is based on the assumption that the speed of the target and incident electrons is much higher than that of the nuclei, so that an electronic transitions take place in a very short time, during which the nuclei remain still at a given internuclear distance. In this paper we are interested on vibrationally excited target and then we limit the calculation to rotationally averaged cross sections only~\cite{Lane}. Moreover, as the final electronic state $\textrm{A}\,^2\Sigma_g^+$ is essentially repulsive, so that its vibrational spectrum form a continuum and the excitation to one of these vibrational levels is followed by dissociation, the final total dissociation cross section implies an integration over the continuum vibrational spectrum. Namely:
\begin{equation}
\sigma_v^{X,A}(\epsilon) = \int_{\epsilon_{th}}^E d\epsilon_{c} \frac{d\sigma_{v,\epsilon_{c}}^{X,A}(\epsilon)}{d\epsilon_{c}} = \frac{\pi}{k^2} \sum_{S,\Lambda,l,l'} g_s \int_{\epsilon_{th}}^E d\epsilon_{c}\left|\left< c \left|T_{l,l'}^{\Lambda,S}(R;k)\right|v\right>\right|^2\,, \label{diss_xsec}
\end{equation}
where $\epsilon$ is the incident electron energy and $k = (2\,m_e\,\epsilon/\hbar^{\/2})^{1/2}$ its momentum. $T_{l,l'}^{\Lambda,S}(R;k)$ is the $R$-depending $T$-matrix expressed in terms of the $l,l'$ partial wave quantum numbers, $\Lambda $ is the group symmetry index and $S$ is the spin state quantum number of the global electron-molecule system. $g_S$ is a spin multiplicity factor, given as $(2S+1)/2(2S_i+1)$, where $2(2S_i+1)$ is the initial total spin. It assumes the values 1/4 and 3/4. $l$ and $l'$ run over the values from 1 to 3 and 1 to 9 respectively, $1\leq\Lambda\leq 16$ for the $D_{2h}$ symmetry group. $\epsilon_{c}$ is the continuum energy and $\epsilon_{th}$ the continuum threshold. Finally, $E$ is the total energy defined by $E=\epsilon + \epsilon_{v}$, where $\epsilon_v$ is the initial vibrational energy eigenvalue. The $T$-matrix has been calculated with the $R$-matrix method for 1000 values of the internuclear distance $R$ in the interval $[1.0, 6.0]$~a.u.  for $0\leq v\leq 15$ and for 1800 $R$-values in the interval $[1.0, 10.0]$~a.u. for $v > 15$. This is done because the continuum wave functions extend to large internuclear distances.

Eq.~(\ref{diss_xsec}) can be cast in an approximate form. Assuming, in fact,
\begin{equation}
\int_{\epsilon_{th}}^E d\epsilon_c \frac{d\sigma_{v,\epsilon_{c}}^{X,A}(\epsilon)}{d\epsilon_{c}}\approx \int_{\epsilon_{th}}^\infty d\epsilon_{c}\frac{d\sigma_{v,\epsilon_{c}}^{X,A}(\epsilon)}{d\epsilon_{c}}\,,
\end{equation}
using the closure property $\int d\epsilon_c\left|c\rangle\langle c\right|= 1$ and applying  then the FC approximation at a particular value of the internuclear distance $R=\bar{R}$, it can be written as:
\begin{equation}
\sigma_{v}^{X,A}(\epsilon)=\frac{\pi}{k^2} \sum_{S,\Lambda,l,l'}g_s\left|T_{l,l'}^{\Lambda,S}(\bar{R};k)\right|^2 \,. \label{fixed-nuclei_xsec}
\end{equation}
This is the approximate form of the total cross section generally used in $R-$matrix calculations, where, usually, $\bar{R}$ is taken at the equilibrium internuclear distance. It will be useful in the discussion of our results in the following sections.

\section{Results \label{sec:results}}
\begin{figure}
\centering
\includegraphics[width=5cm, angle=-90]{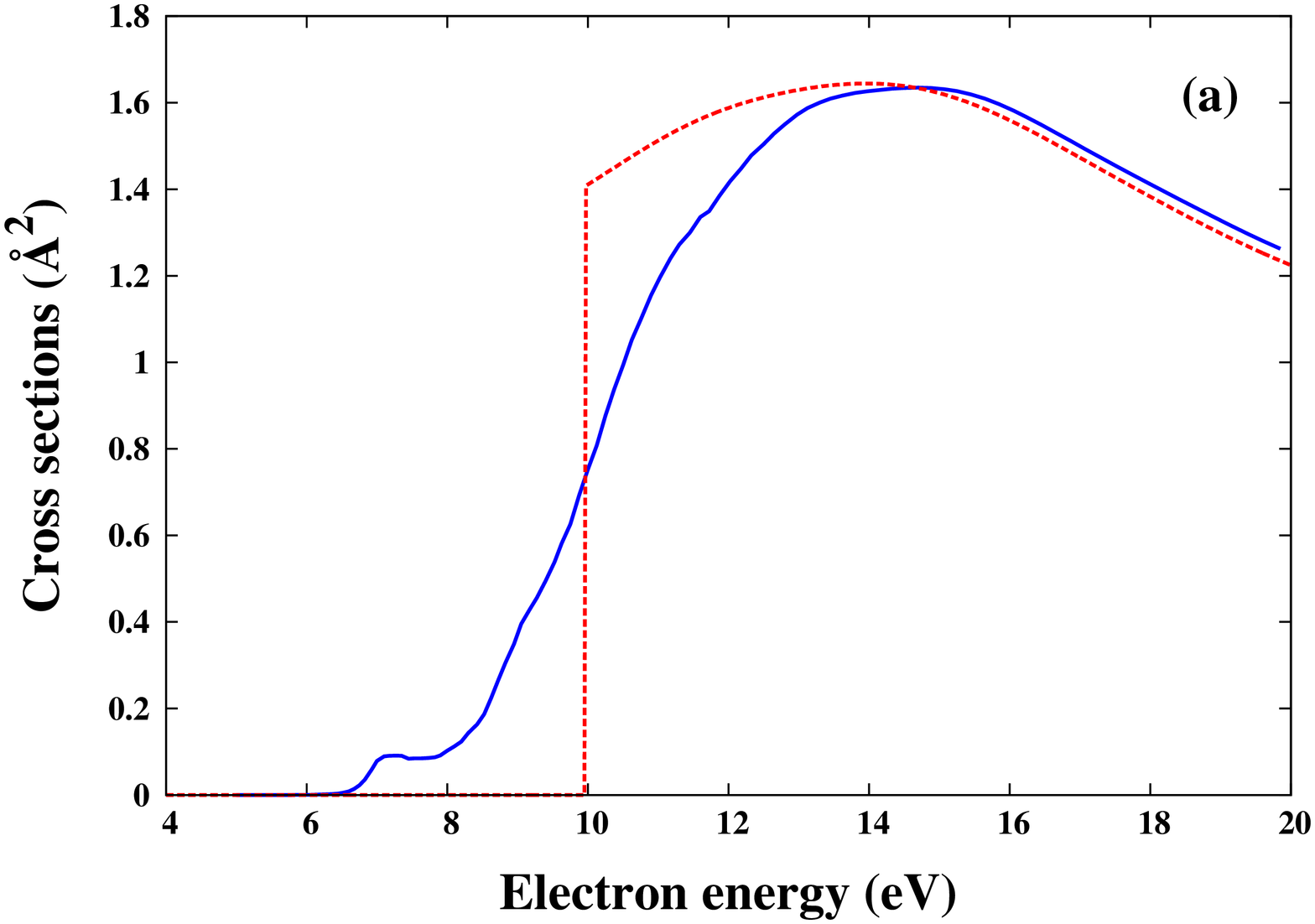}
\includegraphics[width=5cm, angle=-90]{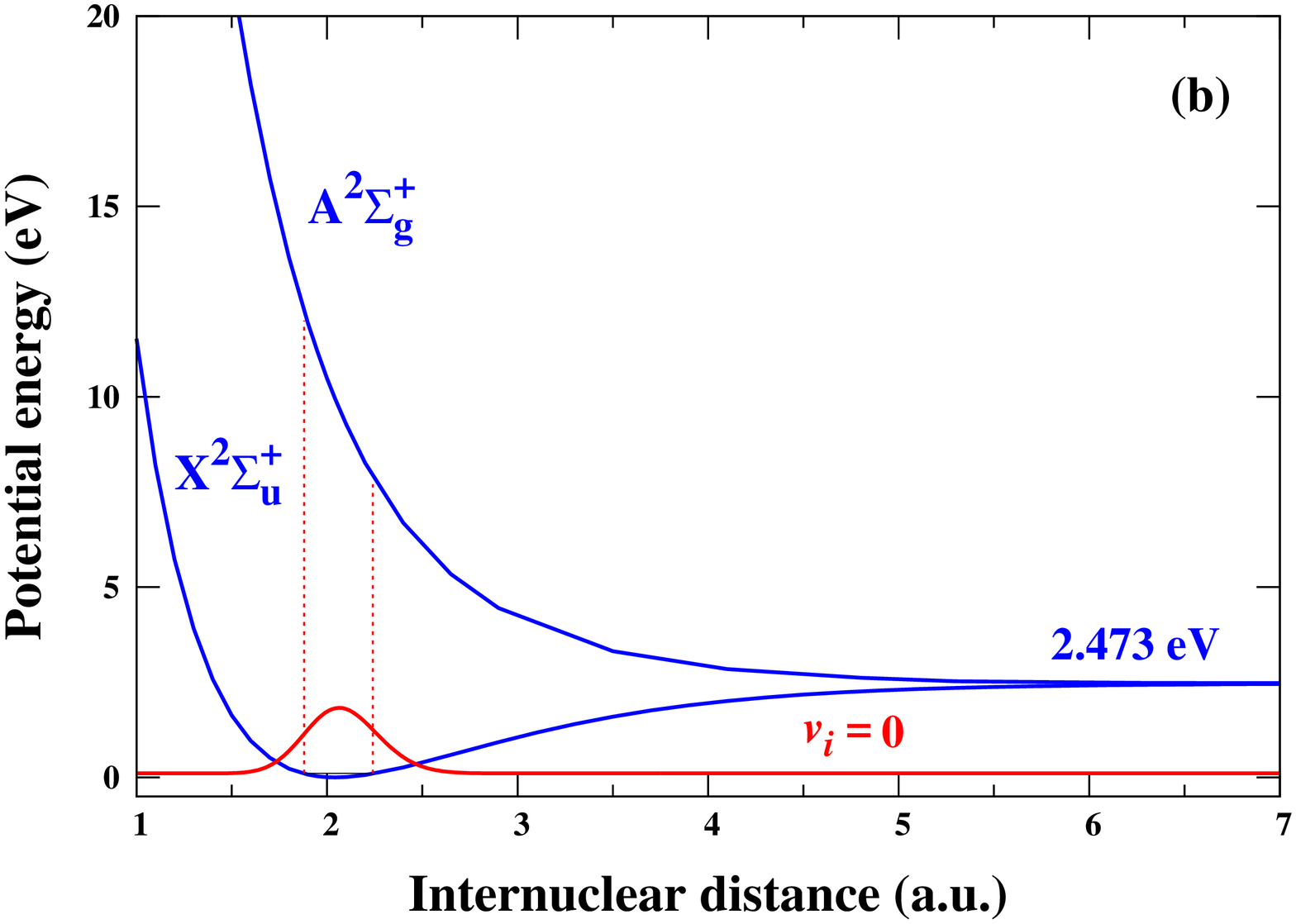}
\caption{\small ($a$) Cross section as a function of the incident electron energy for process (\ref{process}) starting from $v=0$, calculated in the AN (full-blue line) and in the fixed-nuclei (dot-red line) approximations. ($b$) Potential energy curves and $v=0$ wave function (arbitrary units) as a function of the internuclear distance. The two vertical dashed lines, starting from the classical turning points of the zero-level, enclose the FC region. The zero-energy is placed at the bottom of the ground state well. \label{fig:1}}
\end{figure}
Figure~\ref{fig:1}(\textit{a}) shows the cross sections for $v=0$, as a function of the incident electron energy, calculated in the AN approximation by Eq.~(\ref{diss_xsec}) (full-blue line). The cross section curve exhibits a smooth behavior in all the considered range of energies, reaches a maximum at 14--15 eV and then decreases for higher energies. A structure, characterized by a small maximum positioned at about 7 eV, appears in the energy range of 6.5--8 eV.

The general trend of the cross section can be explained by inspecting the behavior of the vibrational wave function for the level $v=0$ in connection with the electronic state potential curves as depicted in Fig.~\ref{fig:1}(\textit{b}). The right panel of the figure shows that the threshold of the process is at 2.369 eV, counted from the $v=0$ level, but the apparent threshold is observed at $\sim$ 6.5 eV (see Fig.~\ref{fig:1}(\textit{a})), corresponding to the transition energy at $R=2.39$~a.u., close to the outer limit of the FC region.

\begin{figure}
\centering
\includegraphics[width=5cm, angle=-90]{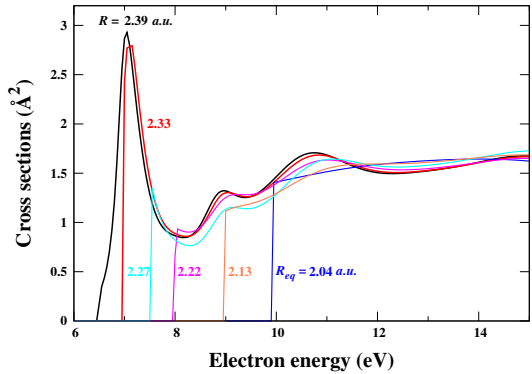}
\caption{\small $R$-matrix cross section calculated at the bond-lengths shown in the plot. $R_{eq}$ denotes the equilibrium internuclear distance. \label{fig:FN}}
\end{figure}

The local maximum at 7 eV is generated by the formation of a resonance occurring in the low energy region. This can be better appreciated in Fig.~\ref{fig:FN}, where the cross sections calculated by Eq.~(\ref{fixed-nuclei_xsec}) for a number of bond-lengths, are shown. The fixed-nuclei cross sections calculated at $R=2.39$~a.u. and $R=2.33$~a.u. exhibits a sharp peak close to 7 eV, while in the other curves, calculated for decreasing internuclear distances, the maximum gets reduced and finally disappears for $R\geq 2.13$~a.u. This is consistent with fact that the threshold of the process, correspondingly, increases and at large collision energies the formation of a resonant state is prevented by the small probability of electron capture. The resonance maximum observed at the two largest bond-lengths in Fig.~\ref{fig:FN}, induces the formation of the 7-eV peak in the AN cross sections, whose magnitude, however, is greatly reduced with respect to the fixed-nuclei cross section. This is due to the fact that the two above resonance bond-length values fall in the right edge of the FC region, where the $v=0$ vibrational wave function is going to vanish (see Fig.~\ref{fig:1}(b)).

\begin{figure}
\centering
\includegraphics[width=5cm, angle=-90]{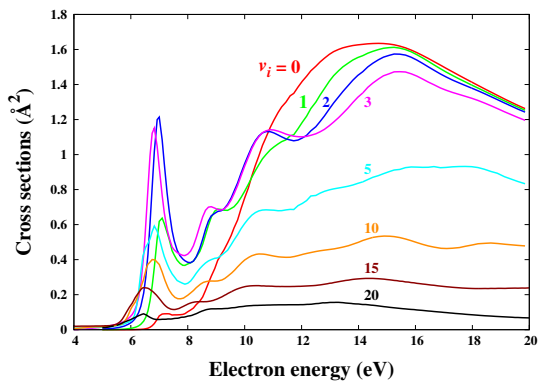}
\caption{\small AN cross sections for processes starting from different He$_2^+ $ vibrational levels $v$.\label{fig:vi}}
\end{figure}

In Figure \ref{fig:vi} the AN cross sections are shown for different vibrational levels. For these cases, the peak maximum increases up to $v=2$, then reduces and its position gets slightly shifted toward lower energies. This behavior is probably due again to the modulation of the vibrational wave function which, for $v> 0$, is peaked at the right edge of the FC region and this could enhance the resonance peak arising in the fixed-nuclei cross sections. For higher levels, instead, the FC region extend far from the resonance bond-lengths so that the wave function maximum no longer overlap with the resonance peak and the AN cross section maximum becomes less pronounced.

At high energies, the AN cross section for $v=0$, in Fig.~\ref{fig:1}(a), strictly follows the fixed-nuclei cross section calculated at the equilibrium internuclear distance, $R_{eq}=2.042$~a.u. (red-dotted curve), showing that for high energies the fixed-nuclei approximation furnishes quite accurate results. For higher vibrational levels, Fig.~\ref{fig:vi} shows a decreasing trend of the cross section for large energies. This is a counter intuitive behavior if one considers that the threshold energies gets smaller with the increasing of $v$. Probably, again the vibrational wave function, whose amplitude decreases for high levels, plays a decisive role in affecting the cross sections. Finally, the oscillations observed in the cross section curves for $v>0$ can be attributed to oscillating behavior of the FC densities also observed in other circumstances \cite{Celiberto_et_al}.

\section{Conclusions \label{sec:conclusions}}
We have calculated the adiabatic nuclei approximation cross sections for electron-impact dissociation of vibrationally excited He$_2^+$ cation. For the case $v=0$, the cross section has a smooth energy behavior, presenting a structure only at the low energy generated by the formation of a resonant state. The intensity of the resonance peak is reduced by the small values of the decaying part of the vibrational wave function in the external side of the FC region, where the resonance arises. For $v>0$ the structure is enhanced, with respect to the previous case, by the corresponding vibrational wave functions which reach their maximum at the outer edge of the FC region. At larger energies, the general decreasing trend of the cross sections is likely determined, once again, by the corresponding behavior of the initial state vibrational wave function amplitude, while the oscillating structures of the cross section curves can be attributed to the influence of the continuum FC density.

\end{document}